\documentclass[11pt,twoside]{article}

\usepackage{asp2006}
\usepackage{epsf}
\usepackage{psfig}
\usepackage{lscape}

\newcommand{\nc}{\newcommand}  
\nc{\w}{$w_2(\theta)$\ } 
\nc{\vev}[1]{\langle #1 \rangle}

\begin{document}
\title{Evidence for dark energy: Cross-correlating SDSS5 with WMAP3}   
\author{A.Cabr\'e$^{1}$,  E.Gazta\~{n}aga$^{1,2}$,
M.Manera$^{1}$,  P.Fosalba$^{1}$ \& F.Castander$^{1}$}   
\affil{\begin{enumerate}
        \setlength{\itemsep}{0pt}
        \setlength{\parskip}{0pt}
        \item \scriptsize{Institut de Ci{\`e}ncies de l'Espai (CSIC/IEEC), Campus UAB, F. de Ci\`encies, Torre C5 par-2, Barcelona 08193, Spain}
	\item \scriptsize{INAOE, Astrof\'{\i}sica, Tonantzintla, Puebla 7200, Mexico}
       \end{enumerate}
}    

\begin{abstract} 
We cross-correlate the third-year WMAP data with galaxy samples extracted from
the SDSS DR5 (SDSS5) covering 16\% of the sky. These measurements confirm a positive cross-correlation, which is well fitted by the
integrated Sachs-Wolfe (ISW) effect for flat LCDM  models 
with a cosmological constant. The combined analysis of different
samples gives $\Omega_\Lambda=0.79-0.83$ (68\% Confidence Level, CL)
and  $\Omega_\Lambda=0.75-0.85$ (95\% CL).  
\end{abstract}

\vspace{-1.cm}

\section{Introduction}
Dark Energy models with late time cosmic acceleration, such as the     
$\Lambda$-dominated CDM model, predict a blueshift in the temperature anisotropies 
of CMB produced by photons coming from last scattering surface that pass through 
matter potentials evolving with time. It is the integrated Sachs-Wolfe (ISW) effect 
which is important at large scales. We can detect the ISW effect through 
the cross-correlation of temperature fluctuations with local tracers 
of the gravitational potential such as galaxies. A positive cross-correlation 
between the 1yr WMAP data (WMAP1) and galaxy samples from the Sloan Digital Sky Survey 
data release 1 (SDSS1) was first found by Fosalba, Gazta\~naga \& Castander (2003), FGC03 from now on, 
and Scranton et al. (2003). 
WMAP1 has also been correlated with the APM galaxies (Fosalba \& Gazta\~naga  2004),
infrared galaxies (Afshordi et al 2004), 
radio galaxies (Nolta et al. 2004), and the
hard X-ray background (Boughn \& Crittenden, 2004). 
Here we want to check if these results can be confirmed to higher significance using
the SDSS data release 5 (SDSS5) which covers $~3$ times the volume of SDSS1. For details see Cabr\'e et al 2006 who has done a similar analysis with DR4.

\section{Results}
In order to trace the changing gravitational potentials we use
galaxies selected from the SDSS5
(Adelman \& McCarthy 2007) (16\% of the sky). We have selected subsamples with
different redshift distributions to check the reliability of the
detection and to probe the evolution of the ISW effect. 
We use a slice with apparent magnitude $r=20-21$ and a selection of high 
redshift galaxies (LRG). For the CMB temperature fluctuations we take the 3rd
year WMAP data (WMAP3).


We define the cross-correlation function as the expectation value of galaxy density    
fluctuations $\delta_G$ and temperature anisotropies    
$\Delta_T=T-T_0$ (in $\mu$K) at two positions $\hat{n}_i$ and $\hat{n}_j$ in the sky.
To estimate the cross-correlation from the pixels maps we average over all
pixels $N_{i,j}$ separated an angle $\theta\pm\Delta\theta$,i.e,  
$w_{TG}(\theta)=\sum_{i,j} \Delta_T({\bf\hat{n}_i}) 
\delta_G({\bf\hat{n}_j})/ N_{i,j}$.

We use $\sigma_8=0.75$ as given by WMAP and obtain $b\sigma_8$ from the
galaxy autocorrelation function. We fit the cosmological models to the 
cross-correlation measurment. Results are shown in fig. 1.



\begin{figure}[h]    
{\centering{
\epsfysize=4.3cm \epsfbox{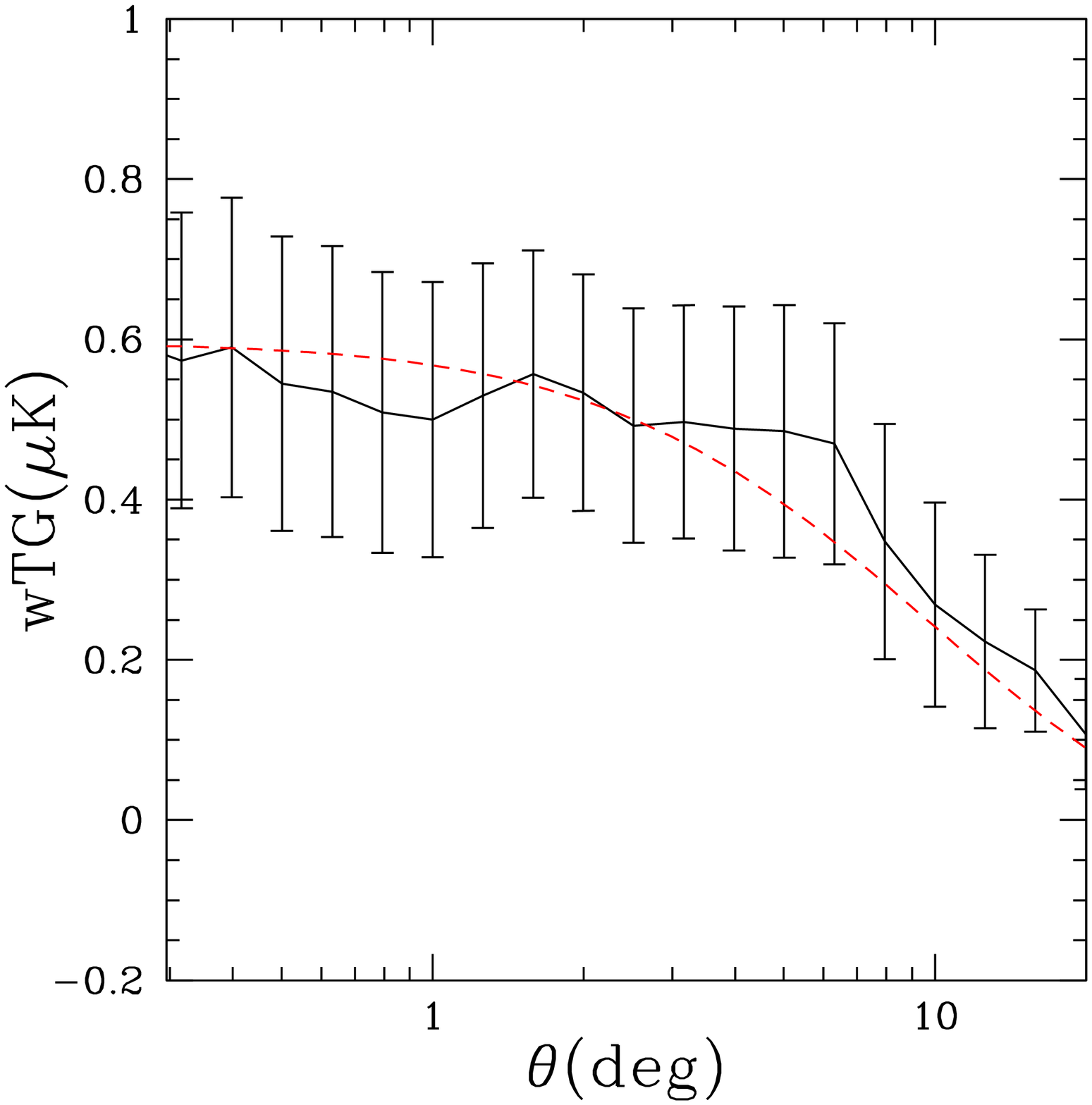} 
\epsfysize=4.3cm \epsfbox{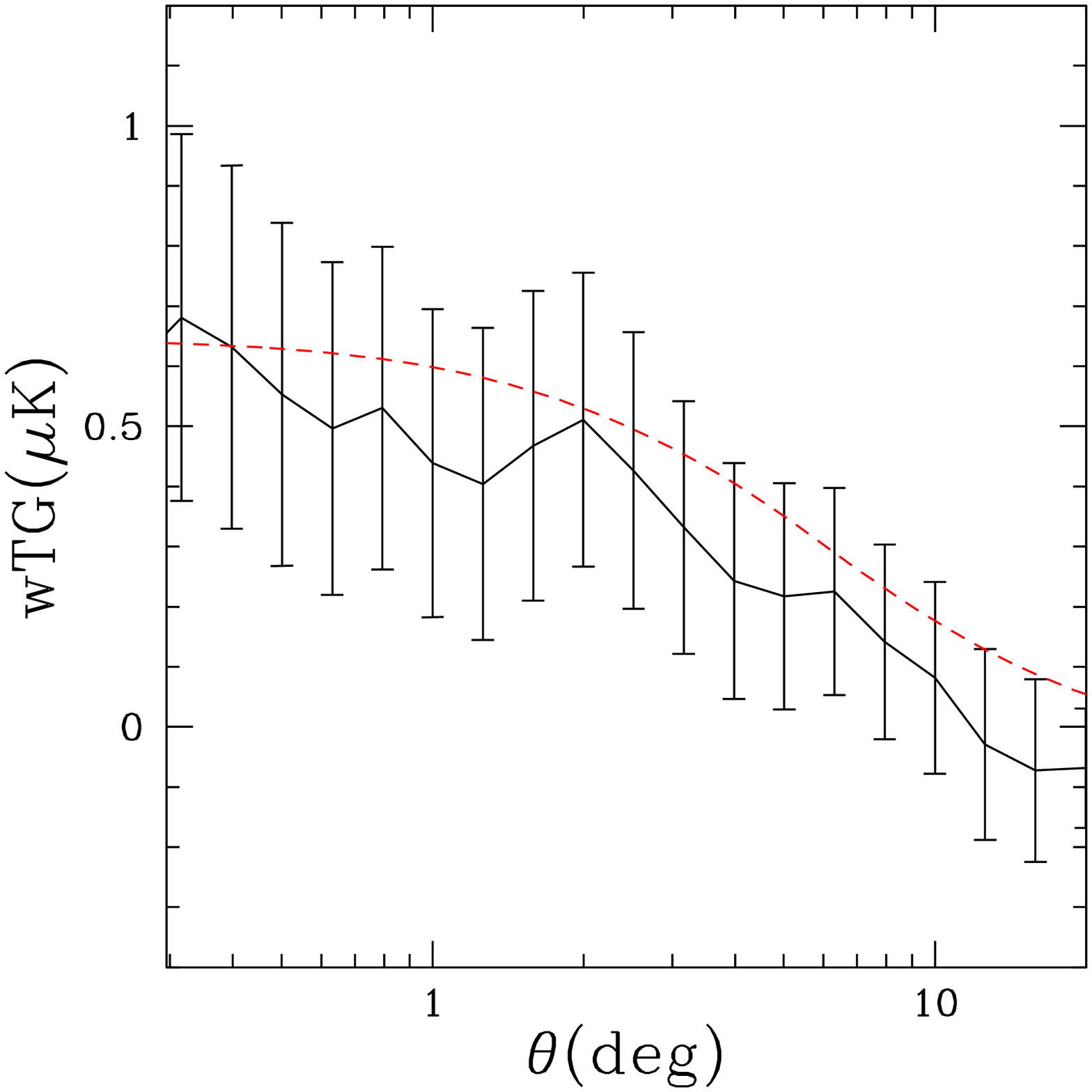}}
\epsfysize=4.3cm \epsfbox{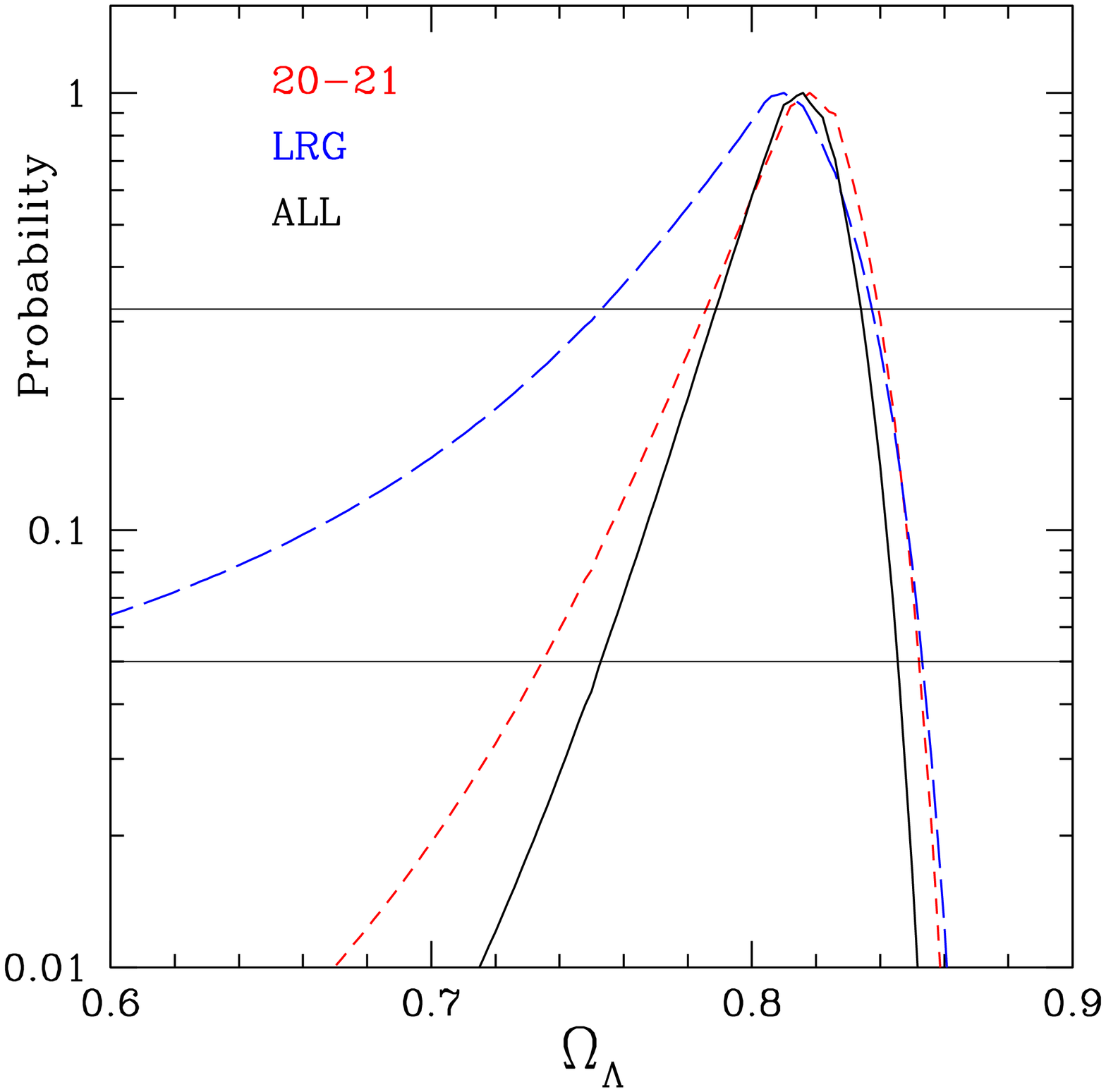}}
\caption{\label{fig:wtg}     
a,b) The continuous line with errorbars shows the
 WMAP3-SDSS5 angular cross-correlation for the
 $r=20-21$ sample and the LRG sample. 
The dashed lines show the $\Lambda CDM$ model with $\Omega_\Lambda =0.81$ ( best overall fit)
scaled to the appropriate bias and projected to each sample redshift.
c) Probability distribution
for $\Omega_\Lambda$ in the $r=20-21$ sample (short-dashed
line), the LRG sample (long-dashed line) and the combined analysis 
(continuous middle curve). The range of $68\%$ and $95\%$ confidence regions
in $\Omega_\Lambda$ are defined by the intersection with the corresponding horizontal lines.}
\end{figure}    

We find that a $\Lambda CDM$ model with $\Omega_\Lambda\simeq 0.81$ successfully 
explains the ISW effect for both samples of galaxies without need
of any further modeling.  The best fit for $\Omega_\Lambda$ for each 
individual sample are very close, as seen in SDSS4-WMAP3.
This is significant and can be understood as a consistency test
for the $\Lambda CDM$ model.

\section*{Acknowledgments}

This work was supported by the European Commission's ALFA-II programme
through its funding of the Latin-american European Network for
Astrophysics and Cosmology (LENAC).


\end{document}